\documentclass[aps,pra,epsfigure,twocolumn,superscriptaddress]{revtex4-1}
\usepackage{dcolumn}    
\usepackage{bm} 
\usepackage{graphicx}
\usepackage{amsmath}    
\usepackage{latexsym}
\usepackage{amsfonts}   
\usepackage{amssymb}
\usepackage{array}      
\usepackage{epsfig}
\usepackage{txfonts}
\usepackage{color}
\usepackage[colorlinks=true,linkcolor=blue,urlcolor=blue,citecolor=blue]{hyperref}
\usepackage{hyperref}
\usepackage{enumerate}
\usepackage{bbold}

\def \w{{\omega}}

\def \ell{{d}}
\def \av#1{{\langle#1\rangle}}

\newcommand{\ket}[1]{\left\vert#1\right\rangle}
\newcommand{\bra}[1]{\left\langle#1\right\vert}

\newcommand{\bka}[1]{\left\langle#1\right\rangle_a}
\newcommand{\bkg}[1]{\left\langle#1\right\rangle_{\Gamma}}
\newcommand{\bkn}[1]{\left\langle#1\right\rangle_{\Nor}}

\newcommand{\Nor}{{\mathcal{N}}}
\newcommand{\Ji}{J_i^{}}

\newcommand{\sutd}{Engineering Product Development Pillar, Singapore University of Technology and Design, 8 Somapah Road, 487372 Singapore}

\newcommand{\majulab}{MajuLab, CNRS-UCA-SU-NUS-NTU International Joint Research Unit, Singapore}

\newcommand{\ntu}{Division of Physics and Applied Physics, School of Physical and Mathematical Sciences, Nanyang Technological University, Singapore}

\begin{document}

\title{Irreversible Work Reduction by Disorder in Many-Body Quantum Systems}

\author{Yuanjian Zheng}
\affiliation{\ntu}

\author{Dario Poletti}
\affiliation{\sutd}
\affiliation{\majulab}

\begin{abstract}  
We study the effect of disorder on work exchange associated to quantum Hamiltonian processes by considering an Ising spin chain in which the strength of coupling between spins are randomly drawn from either Normal or Gamma distributions. The chain is subjected to a quench of the  external transverse field which induces this exchange of work. In particular, we study the irreversible work incurred by a quench as a function of the initial temperature, field strength and magnitude of the disorder. While presence of weak disorder generally increases the irreversible work generated, disorder of sufficient strength can instead reduce it, giving rise to a disorder induced lubrication effect. This reduction of irreversible work depends on the  nature of the distribution considered, and can either arise from acquiring the behavior of an effectively smaller quench for the Normal-distributed spin couplings, or that of effectively single spin dynamics in the case of Gamma-distributed couplings.	 %Depending on the disorder distribution, the reduction of irreversible work can be due to an effective smaller quench(e.g. for the Normal distributed spin couplings) or to an effective single spin dynamics (e.g. for Gamma distributed ones). 
%We relate this to the fact that, with increase in disorder strength, a larger portion of the system evolves with effectively single site dynamics. While this effect is evident in systems with Gamma-distributed couplings, it is less so for Gaussian-distributed ones, where the relative lubrication decreases with the strength of disorder.         
\end{abstract}

\maketitle

\section{Introduction}    

The recent years have seen a growing interest in the study and implementation of miniaturized heat engines that are able to convert the flow of heat into useful work at the nanoscale \cite{BenentiWhitney2017, Gelbwaser2015, Pekola2015}. Such devices have been experimentally realized with much success in setups of various design and working substances that include trapped ions \cite{RossnagelSinger2016, GKVuletic2018}, nitrogen vacancies \cite{KlatzowPoem2017}, single particle or colloidal systems \cite{HugelGaub2002, MartinezRica2015,BlickeBechinger2012}, piezoresistive devices \cite{SteenekenVanBeek2011}, a micro-meter sized piston \cite{QuintoSu2014}, ultracold gases \cite{BrantutGeorges2013}, quantum dots \cite{ThierschmannMolenkamp2015} and even a single spin \cite{PetersonSerra2018}.

As such, much attention has been dedicated to various important aspects of quantum heat engines and the strategies in enhancing their performance. For instance, the use of shortcuts to adiabaticity in improving the work exchanged of finite-time protocols \cite{DelCampoPaternostro2014, DengGong2013, DeffnerDelCampo2014, BeauDelCampo2016, AbahLutz2017} and the ensuing discussion of its associated cost and trade-offs \cite{ZhengPoletti2016b, CampbellDeffner2017, TorrenteguiMuga2017} have been a topic of intense research. More generally, these strategies include eliminating or reducing friction by optimizing the distribution or absolute amount of time spent on the unitary components of thermodynamic cycles \cite{RezekKosloff2009, AcconciaDeffner2015, ZhengPoletti2016, KosloffRezek2017}, geometry of the confining potential \cite{QuanNori2007, Quan2009, ZhengPoletti2014}, particle statistics \cite{GongQuan2014, ZhengPoletti2015, JaramilloDelCampo2016}, and the use of non-thermal baths \cite{AbahLutz2014}. More recently, there has also been significant interest in thermodynamic cycles that employ measurements in place of heat baths \cite{ElouardAuffeves2016, ElouardAuffeves2017, YiKim2017, DingTalkner2018, ElouardJordan2018}.    

On the other hand, the presence of disorder is known to be capable of significantly altering the properties of a system. Most notably, disorder induces Anderson localization \cite{Anderson1958} in non-interacting one-dimensional systems, while in the presence of interaction, a disordered system can exist in two different phases that are commonly referred to as the ergodic or the many-body localized phases \cite{BaskoAltshuler2006, AbaninSerbyn2018}. 

However, despite its relevance to the dynamics of a system, there have been relatively few studies devoted towards understanding the effects of disorder on quantum work statistics. In \cite{AlecceZambrini2015}, it was shown that disorder in the alignment coupling of spins to an external field increases the inner friction incurred from a quench and would thus limit the performance of an engine cycle. We note that more recently, there is also nascent interest in studying the work distribution from a random matrix perspective \cite{LobejkoTalkner2017, ArraisToscano2018,ChenuDelCampo2017,ChenuDelCampo2018} and that an engine cycle based on many-body localization has also been proposed \cite{HelpernRefael2017}. 

Hence, this work aims to extend our understanding of quantum work exchange of Hamiltonian processes in the broader context of disordered many-body systems. To this end, we study an interacting Ising spin chain in which the spin couplings exhibit quenched disorder, in that they are selected independently and at random from a given probability distribution. We evaluate the mean work done and free energy change under application of a time-dependent transverse field, and show that their resulting difference, also known as irreversible work, grows with the magnitude of disorder for weak values, while it is instead significantly suppressed at large magnitudes of the disorder. We show that this sort of lubrication by disorder occurs when the spin couplings are drawn from a distribution such as the Gamma distribution, for which the probability to obtain relatively small values of the spin-couplings grow significantly with the variance of the distribution. Conversely, we show that while the absolute value of the irreversible work similarly decreases for Normal-distributed spin couplings for strong values of disorder, the relative reduction of the irreversible work, in comparison to the mean work done, decreases.
 
This paper is organized as follows: In Sec.\ref{sec:model} we specify the model and disorder distributions considered; in Sec.\ref{sec:results} we show how irreversible work is affected by the magnitude of disorder at different temperatures and in Sec.\ref{sec:conclusions} we draw our conclusions.         
	
\section{Model} \label{sec:model} 
We consider a one-dimensional quantum Ising chain of size $N$ in the presence of a transverse field. The Hamiltonian is given by 
\begin{equation}
H(t)=-\sum^N_{i=1} \left(\Ji \sigma_i^x \sigma_{i+1}^x + g(t) \sigma_i^z \right)    
\label{eq:hamiltonian}
\end{equation}
where $\sigma^x_i$ and $\sigma^z_i$ are the Pauli matrices of the $i-$th site and $\Ji$ is the coupling strength of the spin-spin interactions between site $i$ and $i+1$ which we take to be randomly and independently drawn from a given distribution $P_a(\Ji)$ where $a$ labels the probability distribution which the $\left\{J_i \right\}$ are drawn from. The magnitude of the time-dependent external field is represented by $g(t)$ which is homogeneous in space. The interaction terms $J_i$ are taken to be frozen, i.e. they remain fixed throughout subsequent time evolution once their values have been determined. Periodic boundary condition is also assumed hereafter (i.e $ \sigma^b_{N+1}=\sigma^b_{1}$ for $b=x,y,z$).

In the absence of an external field [$g(t)=0$], this Hamiltonian reduces to the Edwards-Anderson model for spin glasses \cite{EdwardsAnderson1975} and can be solved exactly for its equilibrium properties, while provisions for spatial disorder in $g$ (which in this case would be site dependent) leads to the random field Ising model that has been the subject of intense study \cite{Fisher1992}. More recently, this model is also found to exhibit a dynamical quantum phase transition \cite{HeylKehrein2013, VoskAltman2014} under a finite-time quench of the external field. In the absence of disorder, a quantum phase transition occurs for $|g|=1$, as the system is paramagnetic for $|g|<1$ and ferro- or antiferro-magnetic for $|g|>1$. 

Here our focus is on the work exchanged during a Hamiltonian quench, and thus we consider a single time-dependent external field $g(t)$ that couples uniformly to all spins of the system.
At time $t=0$, the external field is fixed at $g_0^{}$ and the system, with its given disordered spin couplings, is assumed to be in thermal equilibrium at inverse temperature $\beta$. In other words, the density matrix in the energy eigenbasis of the Hamiltonian at $t=0$ is given by 
\begin{equation}
\rho_{m,n}(0)=\delta_{m,n}\;e^{-\beta E_n(g_0^{})}/Z_{g_0^{}} 
\end{equation} 
where  $Z_{g_0^{}}=\sum_n e^{-\beta E_n(g_0^{})}$ is the partition function and $E_n$ are the eigenvalues of the Hamiltonian for $g=g_0^{}$.  
For a general quench protocol, the system is driven by a time-dependent $g_t=g(t)$ to $g(\tau)$, at time $t=\tau$, such that the state reached at the end of the driving is given by 
\begin{equation}
\rho(\tau)=U_{0,\tau} \; \rho(0) \; U_{0,\tau}^{\dagger} 
\end{equation}
where $U_{0,\tau}=e^{-i\hbar \int^{\tau}_{0} H(g(t)) dt}$ is the unitary time evolution operator and $\rho(\tau)$ is in general non-diagonal in the instantaneous energy eigenbasis at time $t=\tau$. For a given disordered configuration of $\left\{J_i\right\}$, each realization of the quench protocol $g(t)$, results in work exchange defined by a two-time energy measurement \cite{TalknerHanggi2007}  
\begin{equation}
w=E_m(g_\tau^{})-E_n(g_0^{})
\end{equation}
such that, the work exchanged for a given protocol $g(t)$, averaged over the disorder distribution $P_a(\Ji)$, is given by 
\begin{equation}
\langle w \rangle_a=\int  d\{J\} p_a(\{J\})  \sum_{mn}   \left[E_m(g_{\tau}^{})-E_n(g_0^{})\right]  P^{\tau}_{m,n} \; \rho_{n,n}(0)         
\end{equation}
where $\{J\} = \{ J_1, J_2, ... , J_N \}$ represents a given spin coupling configuration, and $p_a(\{J\})$ the probability density of selecting said configuration given explicitly by $p_a({J})=\prod^{N}_{i=1} P_a(\Ji)$. $P^{\tau}_{mn}$ is the transition probability from the $n-$th eigenstate of the Hamiltonian at time $t=0$ $\ket{\psi^0_n}$, to the $m-$th eigenstate of the Hamiltonian at $t=\tau$ $\ket{\psi^{\tau}_m}$, given by 
\begin{equation}
P^{\tau}_{m,n}= \vert \bra{\psi_m(g_{\tau}^{})} U_{0,\tau}  \ket{\psi_n(g_{0})} \vert^2. 
\end{equation}
We clarify that in our notation, $\langle \dots \rangle$ implies an average for a particular disorder realization, while $\langle \dots \rangle_a$ represents the average over disorder realizations drawn from the probability distribution $P_a(\Ji)$.  
It is at this point instructive to recall the Jarzynski equality $\langle e^{-\beta w} \rangle = e^{-\beta \Delta F}$ \cite{Jarzynski1997, Tasaki2000, Kurchan2000, CampisiTalkner2010, HanggiTalkner2015}, where $\Delta F$ is the equilibrium change in free energy given by:
\begin{equation}
\Delta F=-\frac{1}{\beta} \log{\frac{Z_{g_{\tau}^{}}}{Z_{g_0^{}}}}. 
\end{equation}
From here, using Jensen's inequality it follows that  $\langle w\rangle\ge \Delta F$ (for each disorder realization), which implies the existence of a non-negative difference between the work done in any process and the change in free energy 
\begin{equation}
\bka{\w_{irr}}=\bka{ w(\tau) } - \bka{ \Delta F }     
\end{equation}
that is the irreversible work \cite{PlastinaZambrini2014} averaged over the disorder distribution $P_a(\Ji)$.              
The behavior and dependence of $\bka{w_{irr}}$ and $\langle \Delta F\rangle_a$ on the nature of $P_a(\Ji)$ thus becomes the focus of investigation in the subsequent sections. 

\subsection{$\Gamma$ and $\Nor$ distributions} 
We focus on two qualitatively different probability distributions: \emph{Gamma} and \emph{Normal} distributions:
\begin{itemize} 
\item the Gamma ($\Gamma$) distribution is 
\begin{equation}
P_{\Gamma}(\Ji)=\frac{1}{\Gamma_0(k)\theta^{\kappa}}J_i^{{\kappa}-1}e^{-{\Ji/\theta}}
\end{equation}
where $\kappa$ and $\theta$ are the shape and scale parameters respectively and $\Gamma_0(x)$ is the gamma function. The mean of $\Ji$ over $\Gamma$ is given by $\langle\Ji\rangle_{\Gamma}=\kappa\theta$, while its variance $\bkg{J_i^2}-\bkg{\Ji}^2=\kappa\theta^2$. 
\item the Normal($\Nor$) distribution is 
\begin{equation}
P_{\Nor}(\Ji) =\frac{1}{\sqrt{2 \pi \theta }}e^{-(\Ji-\bar{J})^2 / 2\theta}. 
\end{equation}
where the disorder averaged strength of the spin couplings is given by $\av{\Ji}_{\Nor}=\bar{J}$ and the variance is $\theta$. 
\end{itemize} 
In the following we consider distributions for which the average strength of spin couplings is $\bar{J}$, which serves as our unit of energy, and is hence set as $\bar{J}=1$. Disorder in this system can thus be seen as fluctuations from the mean spin ferromagnetic coupling strength of $\bka{\Ji}=1$. It follows that the strength of disorder as characterized by the variance for both $\Gamma$ and $\Nor$ distributions is solely parameterized by $\theta$.         

Here, we note that the $\Gamma$ distributed disorder constrains $\Ji$ to be positive such that the chain remains ferromagnetic locally. For a disorder distribution that does not impose this restriction, as the $\Nor$ distribution, local anti-ferromagnetic couplings $(\Ji < 0)$ can arise. 

\section{Results} \label{sec:results}

 In this section, we examine the role of disorder in the work exchanged associated to a unitary process by focusing on two qualitatively different distributions for the disorder realizations. We demonstrate that the choice of distribution plays an important role in the behavior of resulting irreversible work.  

We consider an instantaneous quench of the transverse field from $g_0^{}$ to $g_{\tau}^{}$ at time $t=0$ (or in other words, $\tau=0^+$). Given the presence of a quantum phase transition at $g_c^{}=1$ in the ordered limit of $\theta \to 0$, we consider a quench in the regions $g>1$ and $g<1$ \cite{g1}. In particular, we consider the cases from $g_0^{}=1/4$ to $g_{\tau}^{}=3/4$, and from $g_0^{}=5/4$ to $g_{\tau}^{}=7/4$. In addition, we also consider an intermediate scenario for which $g_0^{}=3/4$ and $g_{\tau}^{}=5/4$ that contains $g_c^{}$. For each scenario we evaluate the ensemble averaged free energy $\bka{\Delta F}$, the instantaneous work $\bka{w}$ and the corresponding irreversible work $\bka{w_{irr}}$ for chains of size $N=8$ over a large range of disorder (variances) $0.1 < \theta < 10 $ and inverse temperature $0.1 < \beta < 10$ utilizing $M=5\times 10^4$ disorder realizations for each combination of $\beta$ and $\theta$ \cite{N6}. 

\begin{figure}
	\includegraphics[width=\columnwidth]{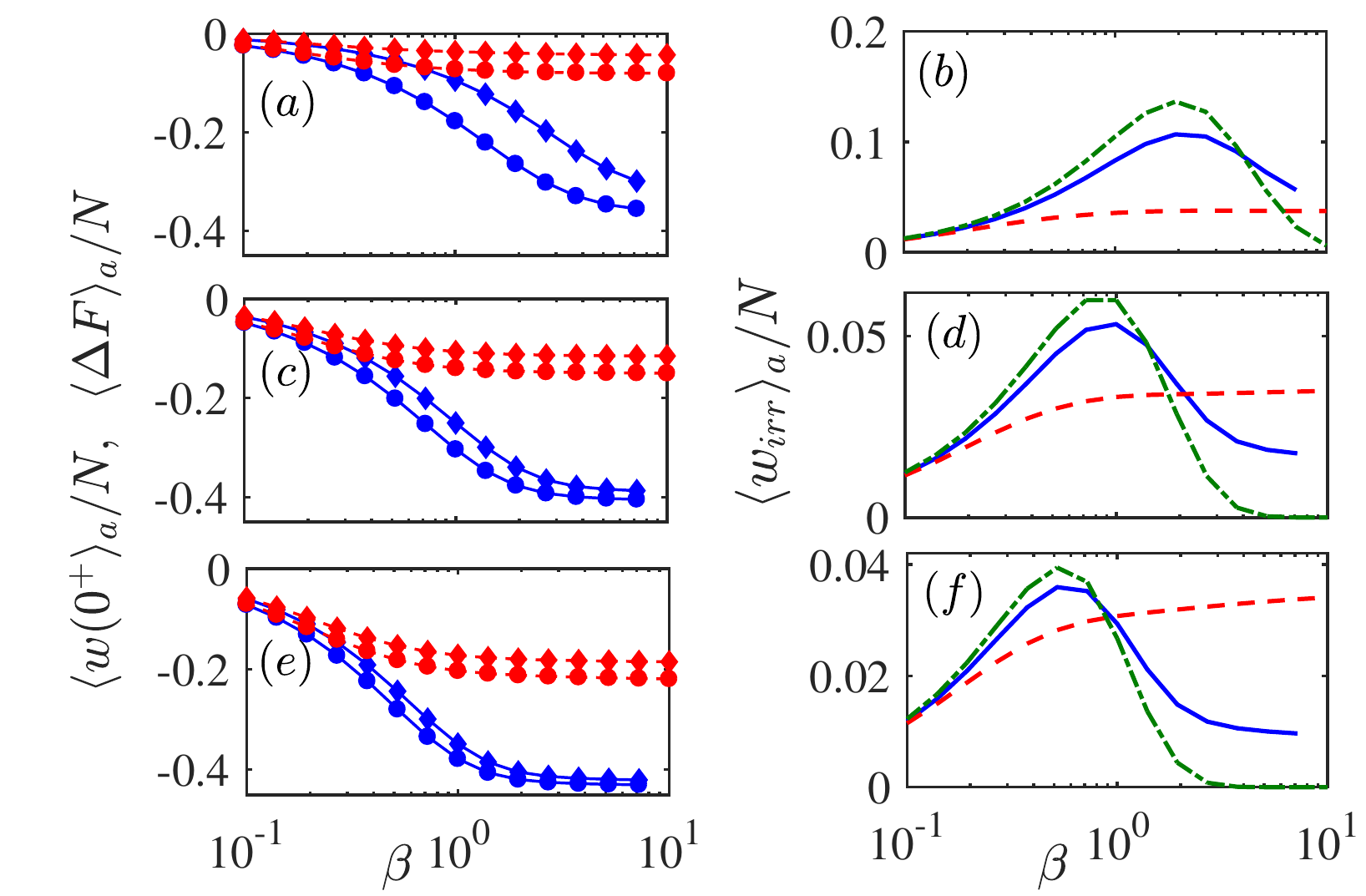}
	\caption{  { \bf [Work statistics for strong disorder] } (a,c,e) $\bka{\Delta F}/N$  (circles) and $\bka{w(0^+)}/N$ (diamonds) for $\Nor -$distributed (red-dashed) and  $\Gamma -$distributed (blue-solid) couplings at $\theta= 10$. 
		(b,d,f) $ \langle w_{irr} \rangle_a / N$ for $\mathcal{N}-$distributed (red-dashed) and $\Gamma -$distributed (blue-solid) couplings  at $\theta= 10$. Analytical approximation of $\bka{w_{irr}}$ (green dot-dashed) given by Eq.(\ref{eq:estimate_wirr}). Quench parameters $g_0^{},g_{\tau}^{}$: (a,b) $1/4 ,3/4$ (c,d) $3/4,5/4$ (e,f) $5/4,7/4$. 	
	}\label{fig:irreversible_work_vs_temperature}
\end{figure}

In Fig.\ref{fig:irreversible_work_vs_temperature} we consider the change in free energy $\bka{\Delta F}$, the instantaneous work $\bka{w(0^+)}$ and the irreversible work $\bka{w_{irr}}$ against inverse temperature $\beta$ at high variance in the coupling distributions ($\theta=10$). %In panels Fig.\ref{fig:irreversible_work_vs_temperature}(a,c,e) the red-dashed line with circles (crosses) represents $\bka{\Delta F}$ ($\bka{w(0^+)}$) for $\Nor -$distributed couplings, while the blue continuous line with circles (crosses) represents $\bka{\Delta F}$ ($\bka{w(0^+)}$) for $\Gamma -$distributed couplings. In panels Fig.\ref{fig:irreversible_work_vs_temperature}(b,d,f) we instead show the irreversible work $\bka{w_{irr}}$ versus $\beta$ for $\Nor-$distributed (red-dashed line) and $\Gamma -$distributed couplings (blue continuous line). 
In panels Fig.\ref{fig:irreversible_work_vs_temperature}(a,c,e) we observe that in the presence of strong disorder, $\bka{w(0^+)}$ seemingly follows the profile of $\bka{\Delta F}$ and both quantities are monotonically decreasing with $\beta$. However, as seen in Fig.\ref{fig:irreversible_work_vs_temperature}(b,d,f), their difference - $\bka{w_{irr}}$, exhibits a qualitatively different behavior as a function of $\beta$. In particular, while all quantities including $\bka{w_{irr}}$ are independent of the chosen disorder distribution in the high temperature (small $\beta$) regime, their behavior departs as $\beta$ increases. Specifically, we see that for the given strength of disorder, $\bka{w_{irr}}$ is non-monotonic with $\beta$ for $\Gamma -$distributed couplings, but is monotonically increasing for $\Nor -$distributed ones. In fact, for sufficiently large disorder, or when the initial state is more local (corresponding to larger values of $g_0^{}$), the irreversible work for $\Gamma -$distributed couplings is lower than that of the $\Nor -$distributed ones. To gain further insights into this difference between the two distributions, we next study the same quantities as a function of $\theta$.

\begin{figure}
	\includegraphics[width=\columnwidth]{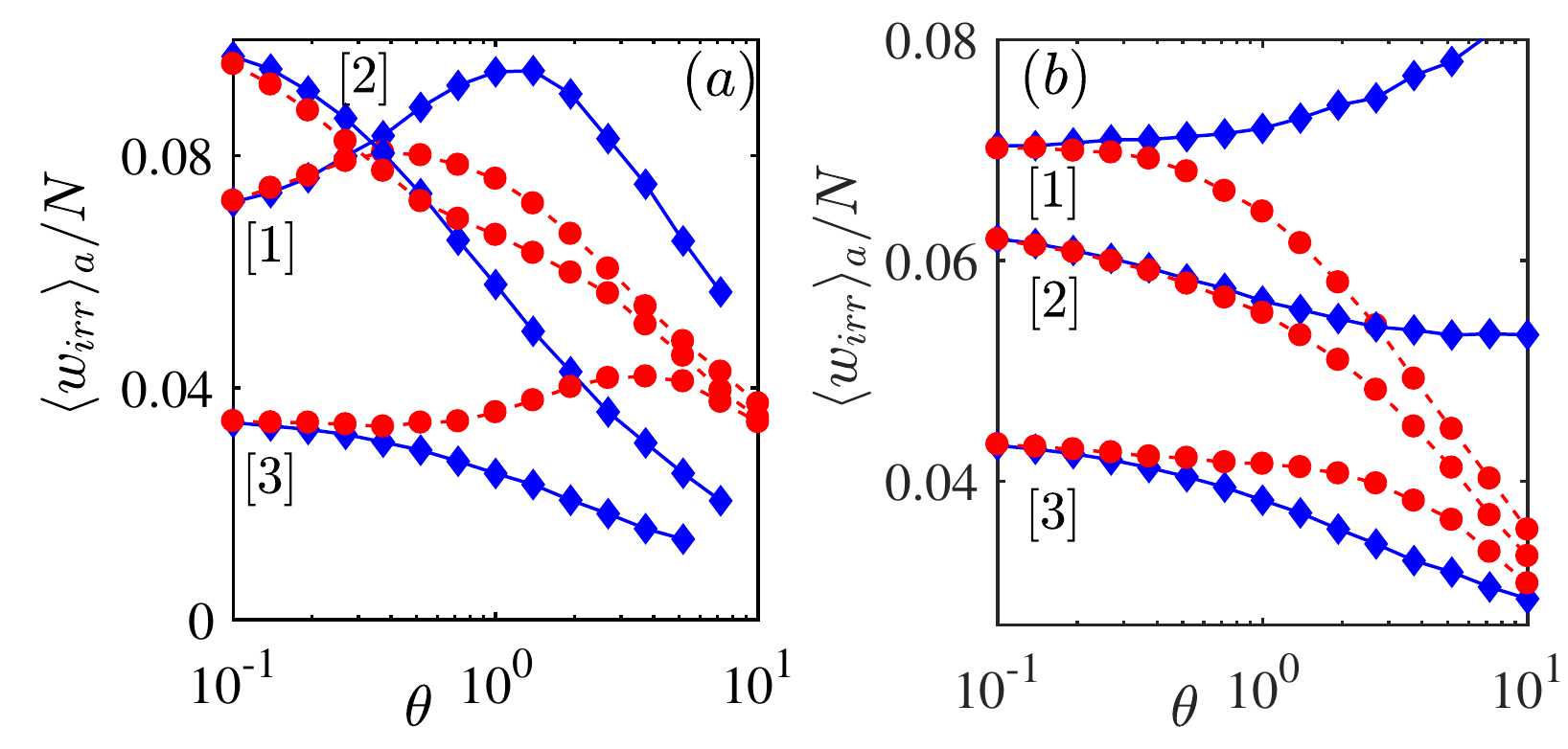}  
	\caption{ {\bf [Irreversible work at moderate and low temperatures]} $\bka{w_{irr}}/N$ for $\mathcal{N}-$distributed (red-dashed) and $\Gamma-$distributed (blue-solid) couplings at (a) $\beta=10$ and (b) $\beta=1$. Quench parameters $g_0^{}, g_{\tau}^{}$, for each set of curves from top to bottom of (a,b) referenced at $\theta=0$ are [1]: $1/4 ,3/4$, [2]: $3/4 ,5/4$ , [3]: $5/4,7/4$.
	}\label{fig:irreversible_work_vs_disorder}
\end{figure}

Fig.\ref{fig:irreversible_work_vs_disorder} shows the irreversible work $\bka{w_{irr}}$ against $\theta$ for the three different quenches considered at (a) low ($\beta=10$), and (b) intermediate ($\beta=1$) temperatures. At both values of $\beta$,  $\bka{w_{irr}}$ appears to converge within the range of disorder considered for $\Nor -$distributed couplings, while remaining essentially unconverged for the $\Gamma -$distributed couplings. However, we note that that the irreversible work, for the $\Gamma -$distributed couplings for quenches that involve larger values of $g$ and strong disorder, is smaller than for $\Nor -$distributed couplings.   
%this converged on value is still larger than that of the $\Gamma$-distributed couplings for quenches that involve larger values of $g$ at similar disorder strengths.               

In short, we find that the presence of disorder can generically reduce the degree of irreversibility in all three quenches examined. However, this reduction is more significant for $\Gamma -$distributed couplings and for more local Hamiltonians. For a deeper understanding of why this is so, we consider the scenario in which the field strength, disorder and temperatures are all taken to be sufficiently large, such that we can approximate the Hamiltonian of the spin chain Eq.(\ref{eq:hamiltonian}) by a local system in which the spins are decoupled from each other such that each spin degree of freedom yields eigenenergies $\pm g$. In this scenario, we can analytically compute both the free energy and the instantaneous work, resulting in an expression for the irreversible work given by 
\begin{equation} 
%w_{irr}  \sim \langle \tilde{w}_{\tau=0}\rangle -\Delta \tilde {F} \geq 0 =-\frac{1}{2} \tanh (\beta g_0^{})+\frac{1}{\beta}\log{ \left(  \frac{\cosh{\beta g_1^{}}}{ \cosh{\beta g_0^{}}}\right)} \geq 0
\bka{w_{irr}}  \sim -(g_1^{}-g_0^{}) \tanh (\beta g_0^{})+\frac{1}{\beta}\log{ \left(  \frac{\cosh{\beta g_1^{}}}{ \cosh{\beta g_0^{}}}\right)}. \label{eq:estimate_wirr}      
\end{equation}
         
In Fig.\ref{fig:irreversible_work_vs_temperature} we see that Eq.(\ref{eq:estimate_wirr}), represented by the green dot-dashed line reproduces qualitatively $\bka{w_{irr}}$  computed for the $\Gamma -$distributed spin couplings, most notably by exhibiting the non-monotonic behavior with $\beta$ that is not present (at these disorder strengths) in the  $\Nor -$distributed chains, and thus becoming significantly smaller at larger $\beta$. 

The fact that the single spin description can be used fairly successfully for $\Gamma -$distributed spin couplings, in contrast to $\Nor -$distributed ones, arises from the inherent difference in the behavior of the two distributions in the strong disorder limit. As discussed earlier, a significant amount of weak coupling terms $\Ji$ arises for the $\Gamma -$distribution for large $\theta$, since it diverges in the limit of $\Ji  \to 0$ for $\theta > 1$. This however, does not occur for the $\mathcal{N}-$distribution since there is no constraint that each $\Ji$ remains positive even as the variance of the distribution increases. As such, the $\Nor$ distribution is thus able to attain an average $\bkn{\Ji}=1$ without incurring a large probability of having weak spin couplings. Consequentially the decoupled spins picture is a much better approximation for the $\Gamma -$distributed spin couplings. 

On the contrary, the probability that $\Ji/g_{\tau}>>1$ increases for the $\Nor$ distribution with increasing $\theta$. This implies that the energy scale of the spin-spin coupling term dominates, and as a result $\bka{w}$, $\bka{\Delta F}$ and $\bka{w_{irr}}$ all asymptotically converges to zero in the limit of $\theta \to \infty$. 

To confirm this understanding on the consequences of the qualitative differences of these distributions, we plot representative probabilities of obtaining either small ($P_a^S$) or large ($P_a^L$) values of $\Ji$.
\begin{equation}
P_a^S=\int_{-\gamma g_0^{}}^{\gamma g_0^{}}P_a(\Ji)d\Ji
\end{equation}
\begin{equation}
P_a^L=\int_{-\infty}^{- g_{\tau}^{}/\gamma}P_a(\Ji)d\Ji + \int^{\infty}_{ g_{\tau}^{}/\gamma}P_a(\Ji)d\Ji
\end{equation}
These correspond to the probabilities that $|\Ji|<\gamma g_0^{}$, (i.e. that the $\Ji$ are small), and that {$|\Ji|> g_{\tau}^{}/\gamma$}, (i.e. that the $\Ji$ are large) respectively. In the following we will consider $\gamma=2$.     

\begin{figure}
	\includegraphics[width=\columnwidth]{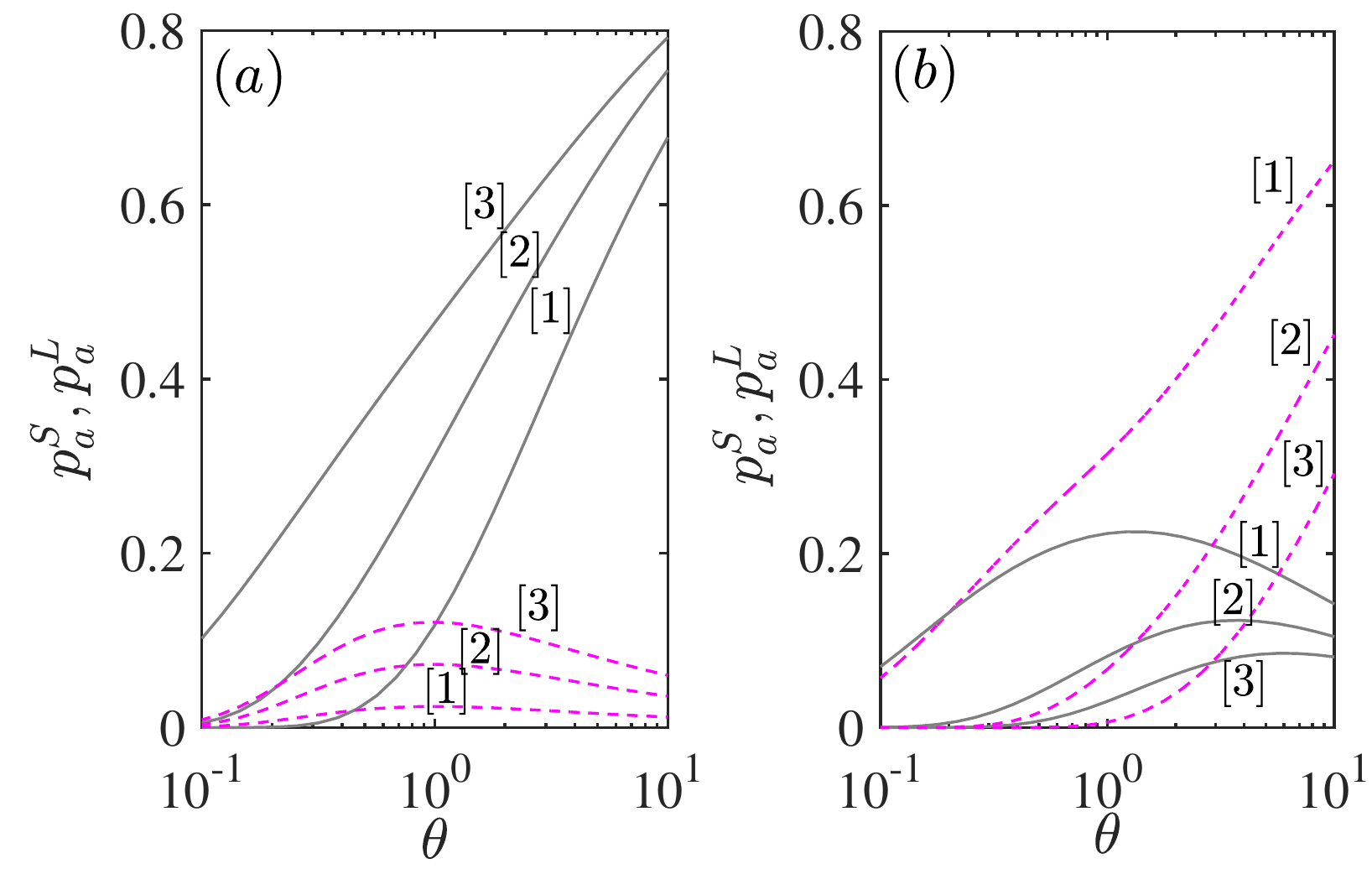}
	\caption{{[\bf Probability of obtaining weak or strong spin couplings]} Probability of obtaining (a) weak $P^S_a $ or (b) strong  $P^L_a$ values of $J_{i}$ for a given strength of the disorder $\theta$ for $\Gamma -$ (gray solid line) and $\mathcal{N}-$ (dashed pink line) distributions, for quench parameters $g_0^{}, g_{\tau}^{}$, [1]: $1/4 ,3/4$, [2]: $3/4 ,5/4$ , [3]: $5/4,7/4$.		
	}	\label{fig:prob_ana}
\end{figure}

These probabilities are plotted in Fig.\ref{fig:prob_ana}(a,b) respectively for $P^S_a$ and $P^L_a$. 
Indeed, Fig.\ref{fig:prob_ana} shows that the occupation of weak values for the $\mathcal{N}$ distribution remains small for the full range of $\theta$ considered. Similarly, not only does $P_{\Gamma}^S$ increases with the variance $\theta$, but it also reaches large values for the maximal $\theta$ considered. On the other hand, this behavior is reversed for the two distributions when the occupation of strong values is considered, thus confirming our intuition of the physical picture. 

\begin{figure}
	\includegraphics[width=\columnwidth]{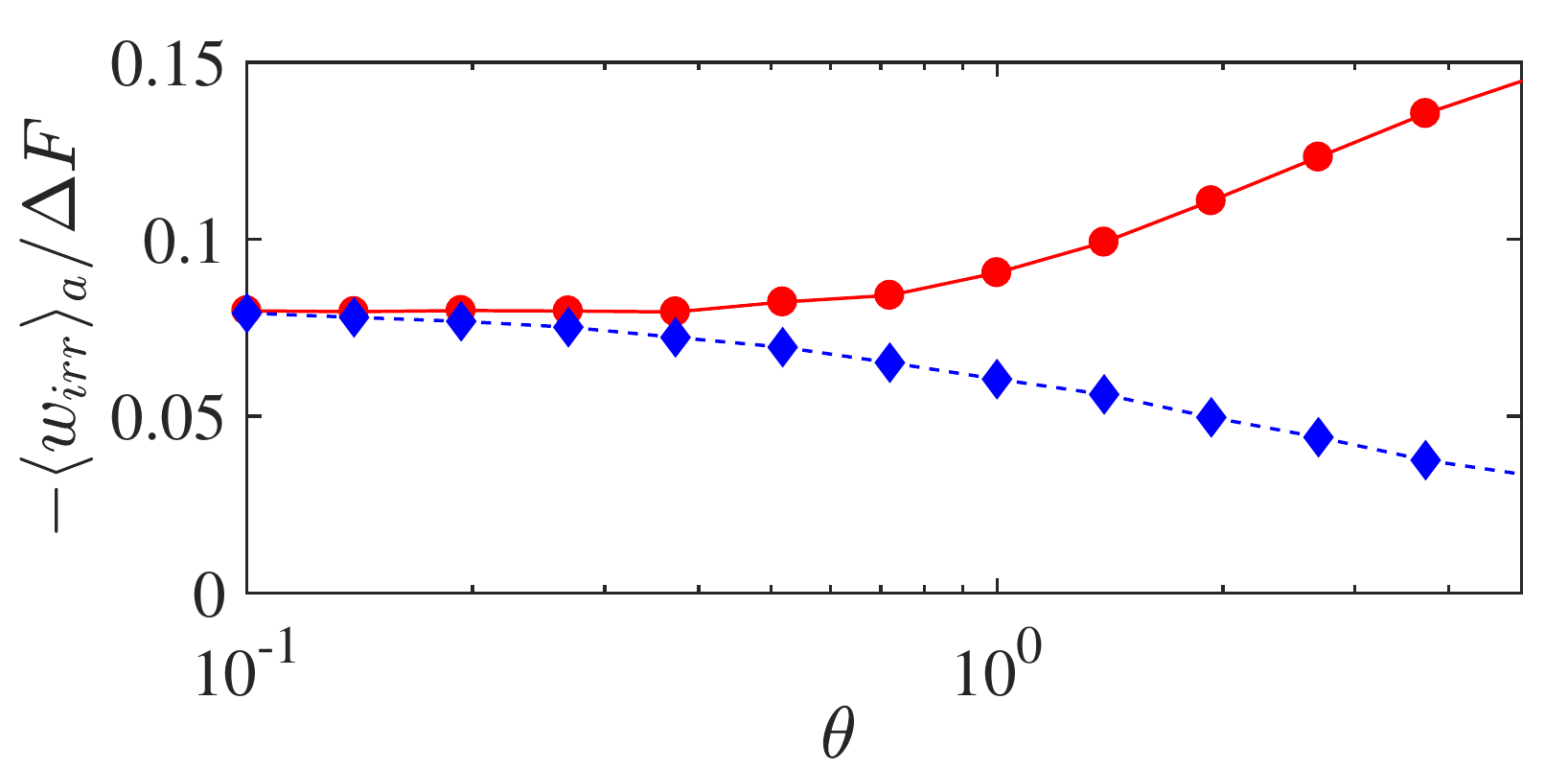}
		\caption{{[\bf Ratio of irreversible work to free energy change]} $-\bka{w_{irr}} / \Delta F$ for $\mathcal{N}-$distributed (red-solid) and $\Gamma -$distributed {(blue-dashed)} couplings at $\beta=10$ for a quench from $g_0^{}=5/4$ to $g_{\tau}^{}=7/4$.    
	}	\label{fig:ratio}
\end{figure}
The behavior of various thermodynamic quantities considered at different temperatures and disorder strength suggests that the correct use of disorder could be a general design principle in minimizing inefficiencies of finite-time quantum processes by reducing the irreversible work incurred. However, as also pointed out earlier, the decrease of $\bka{w_{irr}}$ can also be accompanied by a vanishing $\Delta F$, such that it is unclear from just $\bka{w_{irr}}$ alone if disorder is indeed effectively serving as a lubricant to the dynamics of the system. 

We thus study the ratio $-\bka{w_{irr}} / \Delta F$ as a measure of the effective irreversibility for a wide range of disorder strengths, temperatures and quench regimes for both $\mathcal{N} -$ and $\Gamma -$distributed spin couplings $\Ji$. This ratio is bounded such that $0 < -w_{irr} / \Delta F < 1$, where zero is the limit of perfect lubrication where no irreversible work is generated. 
In Fig.\ref{fig:ratio} we show the ratio of irreversible work to change in free energy, for a quench from $g_0^{}=5/4$ to $g_{\tau}^{}=7/4$ for $\Nor -$ and $\Gamma -$distributed spin couplings as a function of the distribution's variance $\theta$, at low temperature ($\beta=10$). We see that while for the $\Nor -$distributed couplings, this ratio increases with the variance (red-dashed line), it is clearly reduced for the $\Gamma -$distributed (blue- solid line) spin couplings. There is thus a certain lubrication effect of disorder for $\Gamma -$distributed spin couplings.      	  

\section{Conclusions} \label{sec:conclusions} 

In this work we studied how the work exchange of a many-body Ising spin chain under quench of an external transverse field is affected by the presence of quenched disorder. In particular, we considered two qualitatively different disorder distributions - Normal and Gamma, where we identified the essential difference in our context, to be the effective probability of obtaining considerably weaker or stronger values of the spin couplings at large values of the disorder strength.  

We find that while the absolute irreversible work increases with disorder for effectively weak values, it instead decreases at sufficiently strong disorder for both distributions considered. This effect is more pronounced for spin couplings drawn from the Gamma distribution,  where this is the result of an effective single-spin dynamics. In contrast, this effect is due to a lower relative magnitude of the Hamiltonian quench for Normal-distributed couplings. It should be  however stressed that, unlike for Gamma-distributed couplings, the relative lubrication effect decreases for larger disorder magnitudes for Normal-distributed ones.       

It might be of interest, as a further study, to examine the effects of disorder on a complete thermodynamic cycle that would consist of non-unitary processes in which heat is also transferred. The presence of disorder conceivably affects all forms of energy transfer and hence would also affect the overall efficiency and work transfer of the engine. A study of such thermodynamic processes and quantities in the thermodynamic limit might lead to further important insights into strategies for achieving highly efficient quantum heat engines.

D.P. acknowledges fundings from Singapore MOE Academic Research Fund Tier-2 project (Project No. MOE2014-T2-3222-119, with WBS No. R-144-000-350-112) and  by the Air Force Office of Scientific Research under Award No. FA2386-16-1-4041. Y. Z. acknowledges support from the Singapore Ministry of Education Research Fund (Tier 2) MOE2017-T2-1-066 (S). This research was supported in part by the National Science Foundation under Grant No. NSF PHY-1748958


\begin{thebibliography}{100}
\bibitem{g1} For the system size studied, up to $N=8$, we do not observe a qualitatively different behavior when the quench is accross $g=1$.
\bibitem{N6} We have checked our results numerically for $N=6$ and found no qualitative dependence of our results for these system sizes.
\end{thebibliography}
\end{document}